\documentclass[aps,prl,showpacs,preprintnumbers,amsmath,amssymb]{revtex4}
\usepackage{setspace}
\usepackage{graphicx}
\begin{document}

\title {The Classical and Quantum Mechanics of a Particle on a Knot} 

\author{V. V. Sreedhar\footnote{\sl sreedhar@cmi.ac.in}}  

\affiliation{Chennai Mathematical Institute, Plot H1, SIPCOT IT 
Park, Siruseri, Kelambakkam Post, Chennai 603103, India}

\begin{abstract}

A free particle is constrained to move on a knot obtained by winding around 
a putative torus. The classical equations of motion for this system are solved 
in a closed form. The exact energy eigenspectrum, in the thin torus limit,  
is obtained by mapping the time-independent Schr\"{o}dinger equation to the 
Mathieu equation. In the general case, the eigenvalue problem is described
by the Hill equation. Finite-thickness corrections are incorporated 
perturbatively by truncating the Hill equation. Comparisons and contrasts 
between this problem and the well-studied problem of a particle on a circle 
(planar rigid rotor) are performed throughout. 

\end{abstract}


\maketitle


\section{Introduction}

The example of a particle constrained to move along a circle -- the so-called 
planar rigid rotor -- is one of the simplest problems that is discussed in 
text-books of quantum mechanics. The beguiling simplicity of this problem is 
at the heart of many non-trivial ideas that pervade modern physics. For 
understanding many issues like the existence of inequivalent quantizations of 
a given classical system \cite{bal}, the role of topology in the definition 
of the vacuum state in gauge theories \cite{cole}, band structure of solids
\cite{kittel}, generalised spin and statistics of the anyonic type \cite{wilk},
and the study of mathematically interesting algebras of quantum observables 
on spaces with non-trivial topology \cite{floreanini}, the problem of a 
particle on a circle serves as a toy model.  

In this paper, we consider the problem of a particle constrained to move
on a torus knot. Besides adding a new twist to the aforementioned problems,
the present system can be thought of as a double-rotor (analogous to the 
double-pendulum, but without the gravitational field) which is a genuine
non-planar generalization of the planar rotor. 

The paper is organised as follows: In the next section we introduce toroidal 
coordinates in terms of which the constraints which restrict the motion of 
the particle to the torus knot are most naturally incorporated. As a warm-up, 
we then analyse the particle on a circle in toroidal coordinates. This prelude 
allows us to compare and contrast the results of the subsequent sections with
the well-known results for the particle on a circle. The following two
sections deal with the classical and quantum mechanics of a particle 
on a torus knot. In the penultimate section we briefly touch upon the 
possibility of inequivalent quantizations of the particle on a knot.
These will be labelled by two parameters, in contrast to the particle
on a circle. The concluding section summarises and presents an outlook. 

\section{Toroidal Coordinates}

The toroidal coordinates \cite{morse} are denoted by $ 0\leq\eta<\infty,
 ~~ -\pi < \theta \leq \pi , ~~ 0 \leq \phi <2 \pi $. Given a toroidal
surface of major radius $R$ and minor radius $d$, we introduce a 
dimensional parameter $a$, defined by $a^2 = R^2 - d^2$, and a dimensionless
parameter $\eta_0$, defined by $\eta_0 = {\hbox{cosh}}^{-1}(R/d)$.  
The equation $\eta$ = constant, say $\eta_0$, defines a toroidal 
surface. The combination $R/d$ is called the aspect ratio. 
Clearly, larger $\eta_0$ corresponds to smaller thickness of the 
torus. In the limit $\eta_0 \rightarrow \infty$, the torus degenerates
into a limit circle.

The toroidal coordinates are related to the usual Cartesian coordinates 
by the equations 
\begin{equation}
x = {a{\hbox{sinh}}\eta{\hbox{cos}}\phi\over 
({\hbox{cosh}}\eta - {\hbox{cos}} \theta )},~~~
y = {a{\hbox{sinh}}\eta{\hbox
{sin}}\phi\over ({\hbox{cosh}}\eta - {\hbox{cos}} \theta )},~~~ 
z = {a{\hbox{sin}}\theta\over ({\hbox{cosh}}\eta - {\hbox{cos}}
\theta )}. 
\end{equation}
The metric coefficients are given by the equations
\begin{equation}
h_1 = h_2 = {a\over 
({\hbox{cosh}}\eta - {\hbox{cos}}\theta)},~~~~ h_3 = h_1{\hbox{sinh}}\eta 
\end{equation}
and the volume element is
\begin{equation}
dV = {a^3{\hbox{sinh}}\eta \over ({\hbox{cosh}}
\eta - {\hbox{cos}}\theta)^3} 
\end{equation}
With the help of these basic relations, it is straightforward to rewrite 
well-known Cartesian expressions in toroidal coordinates. 

\subsection{A Particle Constrained to Move on a Circle}

The Lagrangian for a free particle of mass $m$ in Cartesian coordinates 
$(x,y,z)$ is 
\begin{equation}
L = {m\over 2}(\dot x^2 + \dot y^2 + \dot z^2)
\end{equation}
In the above expression, and henceforth, an overdot refers to a time 
derivative. After some algebra, this expression can be rewritten in 
toroidal coordinates as 
\begin{equation}
L = {m\over 2}a^2{(\dot\eta^2 + \dot\theta^2 + {\hbox{sinh}}^2\eta~\dot\phi^2)
\over ({\hbox{cosh}}\eta - {\hbox{cos}}\theta )^2} 
\end{equation}
To restrict the motion of the particle to lie on a circle in the $xy$ plane, 
we impose the constraints $\eta = \eta_0$, a constant, and $\theta = 
\theta_0$, another constant. The Lagrangian then takes the form 
\begin{equation}
L = {ma^2\over 2}{{\hbox{sinh}^2\eta_0} \dot\phi^2\over 
({\hbox{cosh}}\eta_0 - {\hbox{cos}}\theta_0 )^2} 
\end{equation}
The Euler-Lagrange equation
\begin{equation}
{d\over dt}({\partial L\over \partial\dot\phi}) = {\partial L\over\partial\phi} 
\end{equation}
then yields, as expected, 
\begin{equation}
\ddot\phi = 0\Rightarrow \phi (t) = \omega t + \phi_0
\end{equation}
where $\omega$ is a constant and has the physical interpretation of 
frequency, and $\phi_0$ is a constant of integration which specifies 
the position of the particle on the circle at time $t =0$ -- similar
to plane polar coordinates.  

Defining a rescaled mass $M = m{{\hbox{sinh}^2\eta_0}\over 
({\hbox{cosh}}\eta_0 - {\hbox{cos}}\theta_0 )^2}$, we get the Hamiltonian 
$H = {p_\phi^2\over 2M a^2}$ with the momentum canonically conjugate to 
$\phi$ being given by  $p_\phi = Ma^2\dot\phi$ as usual. Using this to set 
up the Schr\"{o}dinger equation and solving it, we get, for the eigenvalues 
and the normalised eigenfunctions respectively,  
\begin{equation}
E_n = {n^2 \hbar^2\over 2M a^2}, ~~~~ 
\psi_n (\phi ) =  {1\over \sqrt {2\pi }}e^{\pm in\phi} 
~~~~n= 0, 1, 2, \cdots 
\end{equation}
For large $\eta_0$, the thickness of the putative torus decreases and 
$M\rightarrow m$: we approach the well-known expressions in plane polar 
coordinates. 

Interestingly, it is also possible to get a particle to move
on a circle by imposing the constraints $\eta = \eta_0$, a 
constant, and $\phi = \phi_0$, another constant. This however 
results in a more complicated Lagrangian {\it viz.}
\begin{equation}
L = {ma^2\over 2} {\dot\theta^2\over 
({\hbox{cosh}}\eta_0 - {\hbox{cos}}\theta )^2} 
\end{equation}
The resulting Euler-Lagrange equation is
\begin{equation}
\ddot\theta ({\hbox{cosh}}\eta_0 - {\hbox{cos}}\theta) = -{\hbox{sin}}\theta~ 
\dot\theta^2 
\end{equation}
which can be re-written as
\begin{equation*}
{d\over dt}[\dot\theta ({\hbox{cosh}}\eta_0 - {\hbox{cos}}\theta)] = 0
\end{equation*}
and readily integrated to yield 
\begin{equation}
\dot\theta ({\hbox{cosh}}\eta_0 - {\hbox{cos}}\theta) = \kappa 
\end{equation}
$\kappa$ being a constant. Thus the solution is reduced to quadratures. Thanks 
to the presence of the factor $({\hbox{cosh}}\eta - {\hbox{cos}}\theta)$, the 
solution is not as simple as the one in plane polar coordinates.  The 
Hamiltonian can be obtained in a straightforward manner and is given by
\begin{equation}
H = {p_\theta^2\over 2m a^2} ({\hbox{cosh}}\eta_0 - {\hbox{cos}}\theta)^2
\end{equation}
The presence of the $\theta$-dependent multiplicative factor is portentous
of additional complications that arise when we make a transition to quantum 
mechanics. In particular, the fact that the conjugate operators $p_\theta$ 
and $\theta$ do not commute requires us to perform an operator-ordering of 
the classical Hamiltonian. 

The above analysis shows that while toroidal coordinates are ideally 
suited to consider the motion of a particle on a circle in the $xy$-plane,
they are more cumbersome when it comes to handling paths which stray 
from the $xy$-plane. Since a knot is intrinsically non-planar, we
should be prepared to confront the attendant complications. It should be 
mentioned, however, that these complications would also be present in other 
coordinate systems. We choose to work with toroidal coordinates because of 
their suitability in imposing the constraints that define a torus knot. 

\section{Classical Mechanics of a Particle on a Knot}

As already mentioned, the constraint $\eta = \eta_0$ defines a toroidal 
surface. A $(p,q)$ torus knot can be obtained by considering a closed path 
that loops $p$ times around one of the cycles of a torus while looping 
around the other cycle $q$ times, $p,q$ being relatively prime integers. 
The desired property can be enforced by imposing the   
constraint: $p\theta + q\phi = 0$. It is easy to check that 
$\theta \rightarrow \theta + 2\pi q \Rightarrow \phi \rightarrow \phi - 
2\pi p$ {\it i.e.} as we complete $q$ cycles in the $\theta$ direction, 
we are forced to complete $p$ cycles in the $\phi$ direction -- as required. 
Imposing the above two constraints on equation (5), we get the Lagrangian 
for a particle on a torus knot to be
\begin{equation}
L = {M\over 2} f(\phi ) \dot\phi^2 
\end{equation} 
where
\begin{equation}
f(\phi ) = {a^2\over (\gamma - {\hbox{cos}}\alpha\phi )^2} ~~~{\hbox{and}}~~~
M = m (\alpha^2 + \beta^2) 
\end{equation} 
with 
\begin{equation}
\alpha = -{q/p}, ~~~\beta = {\hbox{sinh}}\eta_0, ~~~~ 
\gamma = {\hbox{cosh}}\eta_0
\end{equation} 
The main difference between the Lagrangian in (14) and the one for a particle
on a circle viz. equation (6), lies in the appearance of the $\phi$-dependent 
factor $f(\phi)$ in the Lagrangian which contains the information about 
the non-trivial embedding of the knot in three dimensions. 

The Euler-Lagrange equation is given by
\begin{equation}
f(\phi)\ddot\phi + {1\over 2} f^{'}(\phi) \dot\phi^2 = 0
\end{equation}
where the prime denotes a derivative of the function $f$ with respect to
its argument $\phi$. Now, using the above equation of motion, it is 
straightforward to show that  
\begin{equation}
{d\over dt}[\sqrt f \dot\phi ] = 0 \Rightarrow \sqrt f \dot \phi = {\cal A}
\Rightarrow \dot\phi = {{\cal A}\over a}(\gamma - {\hbox{cos}}\alpha \phi )
\end{equation}
where ${\cal A}$ is a constant. Noting that $(1 -\gamma^2) < 0$, the latter
equation can be integrated to get  
\begin{equation}
\phi (t) = {1\over\alpha} {\hbox{tan}}^{-1} \bigl[ \sqrt {\gamma - 1
\over \gamma + 1} {\hbox{tan}}({{\cal A}\alpha\beta t\over 2a})\bigr] 
\end{equation}
In the limit $\eta_0\rightarrow \infty$, the right hand side 
is linear in $t$, as expected for a particle on a circle.

The momentum $p_\phi$ canonically conjugate to $\phi$ and the Hamiltonian $H$
can be easily worked out and are given by the following expressions
\begin{equation}
p_\phi = Mf(\phi )\dot\phi, ~~~~~~H = {p_\phi^2\over 2Mf(\phi )}  
\end{equation}

\section{Quantum Mechanics of a Particle on a Knot} 

In principle, once a Hamiltonian is given, it is a straightforward 
exercise to write down the Schr\"{o}dinger equation. In the present case,
the classical Hamiltonian involves a term which mixes the coordinate
and the canonically conjugate momentum. Since these canonical pairs will
be elevated to the level of operators in the quantum theory, we need 
to prescribe an ordering for the operator products. We choose the 
so-called Weyl ordering which symmetrises the product as follows:
\begin{equation}
H = {1\over 6M} [{1\over f}p_\phi^2 + p_\phi{1\over f}p_\phi + p_\phi^2
{1\over f}] 
\end{equation}
In the above equation, and in what follows, we refrain from putting hats,
but it should be remembered that both $\phi$ and $p_\phi$ are operators 
which obey the canonical commutation relations {\it viz.} $[\phi, 
p_\phi ]_- = i\hbar$. Pulling all the momentum terms to the extreme
right in preparation to make them act on a wavefunction, we get 
\begin{equation}
H = {1\over 2M} [({1\over f}) p^2 - (i\hbar)({1\over f})^{'}p - 
{\hbar^2\over 3}({1\over f})^{''} ]
\end{equation}
In the above form, the Hamiltonian is tailor-made for constructing the 
time-independent Schr\"{o}dinger equation, with the usual prescription for 
replacing the momentum by the corresponding differential operator. The 
resulting Schr\"{o}dinger equation is 
\begin{equation}
[-{\hbar^2\over 2M}({1\over f}){d^2\over d\phi^2} - {\hbar^2\over 2M}({1\over 
f})^{'}{d\over d\phi} - {\hbar^2\over 6M}({1\over f})^{''}] \psi = E\psi 
\end{equation}
The first derivative in $\phi$ can be eliminated by the well-known trick 
of substituting $\psi = \chi\Sigma$ in the above equation and getting rid 
of terms proportional to $d\Sigma/d\phi$ by choosing $\chi$ appropriately. 
This yields for $\chi$,
\begin{equation}
\chi \propto \sqrt f
\end{equation}
Substituting this in (23) and collecting the remaining terms, the 
Schr\"{o}dinger equation reduces to the following equation for $\Sigma$
\begin{equation}
[{d^2\over d\phi^2} + V(\phi)]\Sigma = 0
\end{equation}
where the `potential' $V$ is defined by
\begin{equation}
V(\phi ) = [{2f^{''}f - f^{'2}\over 12 f^2} + {2ME f\over \hbar^2}]
\end{equation}
$V$ is an even function of $\phi$. Substituting for $f$ from (15), we 
get after some algebra, 
\begin{equation}
V = {2MEa^2/\hbar^2 + \alpha^2/2 - \alpha^2\gamma {\hbox{cos}}\alpha\phi/3 
- \alpha^2 {\hbox{cos}}2\alpha\phi/6
\over (\gamma - {\hbox{cos}}\alpha\phi)^2 } 
\end{equation}
Since $V(\phi)$ is a periodic function, the above potential can be expanded
in a Fourier series and equation (25) gets identified with the Hill 
differential equation \cite{whittaker}.
\subsection{The Thin-Torus Approximation}

As already mentioned, large values of $\eta_0$ and hence large values of 
$\gamma$, correspond to a thin torus around which the particle's trajectory 
winds. In this limit, we can restrict to terms of the order of $1/\gamma$. Then
$\beta^2 \sim \gamma^2$, hence $M\rightarrow m$, and equation (27) 
simplifies to 
\begin{equation}
V = {\alpha^2\over 4}\lambda  - {\alpha^2\over 3\gamma}{\hbox{cos}}\alpha\phi
\end{equation}
where 
\begin{equation}
\lambda = {8mE a^2 \over\hbar^2\alpha^2}
\end{equation}
Equation (25) now takes the form 
\begin{equation}
[{d^2\over d\phi^2} + {\alpha^2\over 4}\lambda - {\alpha^2\over 3\gamma}
{\hbox{cos}}\alpha \phi ]\Sigma = 0
\end{equation}
Changing variables such that $\alpha\phi = 2z$, the above equation becomes 
\begin{equation}
[{d^2\over dz^2} + \lambda - {4\over 3\gamma}{\hbox{cos}}2z ]\Sigma = 0
\end{equation}
which is immediately recognised to be the well-known Mathieu equation 
\cite{mclach}. 

The solutions $\Sigma$ of the Mathieu equation with the required periodicity 
are given by the Mathieu functions of fractional order $\nu$ {\it viz.}
\begin{equation}
{\hbox{ce}}_\nu(z,\gamma) = {\hbox{cos}}\nu z - {1\over 6\gamma}\bigr[ 
{{\hbox{cos}}(\nu + 2)z\over (\nu +1) }  
- {{\hbox{cos}}(\nu - 2)z\over (\nu -1) }\bigl]\cdots   
\end{equation} 
\begin{equation}
{\hbox{se}}_\nu(z,\gamma) = {\hbox{sin}}\nu z - {1\over 6\gamma}\bigr[ 
{{\hbox{sin}}(\nu + 2)z\over (\nu +1) }  
- {{\hbox{sin}}(\nu - 2)z\over (\nu -1) } \bigr]\cdots  
\end{equation} 
The complete solution with two arbitrary coefficients $A$ and $B$ is given by
\begin{equation}
\Sigma = A{\hbox{se}}_\nu(z,\gamma) + B {\hbox{ce}}_\nu(z,\gamma) 
\end{equation}

Setting $\nu = {2n\over q}$ where $n$ is an integer, we see that the above 
functions have a periodicity $q\pi$ in $z$, which translates into the 
required periodicity $2p\pi $ in $\phi$. 

The condition relating $\lambda$ to $\nu$ is given by
\begin{equation}
\lambda = \nu^2 + \boxed{2\over 9\gamma^2 (\nu^2 -1)} \cdots
\end{equation} 
Since we are restricting our attention to $1/\gamma$ order, the boxed
terms can be neglected. The allowed values of $\lambda$ follow by setting 
$\nu = {2n\over q}$. These values of $\lambda$, in conjunction 
with equation (29), determine the energy eigenvalues to be 
\begin{equation}
E_n = {n^2\hbar^2\alpha^2\over 2ma^2 q^2} 
\end{equation}
Before proceeding further, it is worth recalling that the complete solution
of equation (23) that we are trying to solve is given by $\psi = \chi\Sigma$ 
with $\chi\propto \sqrt f$. The complete solutions for the (unnormalised)
eigenfunctions $\psi$ with the correct boundary conditions are therefore
given by 
\begin{widetext}
\begin{equation}
\psi^{(n)}_+(\phi) = {a\over (\gamma - \boxed{{\hbox{cos}}\alpha\phi}~)}
\times \{
{\hbox{cos}}(n\alpha\phi /q) ~ \boxed{-{1\over 6\gamma}\bigr[{{\hbox{cos}}
((n + q)\alpha\phi /q)\over (2n/q +1) }  - {{\hbox{cos}}
((n -q)\alpha \phi /q\over 
(2n/q  -1) }\bigl]\cdots}\}   
\end{equation} 
\begin{equation}
\psi^{(n)}_-(\phi) = {a\over (\gamma - \boxed{{\hbox{sin}}\alpha\phi}~)}
\times \{
{\hbox{sin}}(n\alpha\phi /q) ~ \boxed{-{1\over 6\gamma}\bigr[{{\hbox{sin}}
((n + q)\alpha\phi /q)\over (2n/q +1) }  - {{\hbox{sin}}
((n -q)\alpha \phi /q\over 
(2n/q  -1) }\bigl]\cdots}\}   
\end{equation} 
\end{widetext}
where we have used $2z = \alpha\phi$. Further, since we retain only terms of 
order $1/\gamma$, the boxed terms in equations (37) and (38) can be neglected. 

In passing, we mention that the two independent solutions (32) and (33) 
can be combined into a single equation given by \cite{mclach} 
\begin{equation}
\Sigma = e^{i\nu z}u
\end{equation}
where 
\begin{equation}
u = {\hbox{sin}}(z-\sigma) + a_3 {\hbox{cos}}(3z - \sigma ) + 
b_3{\hbox{sin}}(3z -\sigma) + a_5{\hbox{cos}}(5z-\sigma) + b_5{\hbox{sin}}
(5z-\sigma) + \cdots  
\end{equation}
where $\sigma$ is a new parameter such that $\sigma = \pi/2$ yields the 
solution (32) and $\sigma =0$ yields the solution (33). In the above
the coefficients $a,b$ are determined in terms of $\gamma$ and $\sigma$.
To order $1/\gamma$ that we are interested in, only $b_3 = -{1\over 12\gamma}$,
is non-zero. This succinct way of writing the general solution will be 
particularly useful in incorporating finite-thickness corrections.  

It may be noted that for $q=1,~p=-1$, and hence $\alpha = -1$, the above 
results~ (35-38)~reduce to the well-known results for a particle on a circle. 
For a (2,3) torus knot, namely, the trefoil, $p = 2$, and the eigenfunctions 
have a period $4\pi$. The general solutions of Mathieu equations with a period 
$4\pi$ were first worked out by Lars Onsager in 1935 in his dissertation for 
a PhD at Yale \cite{lars}. While it is gratifying to note this, it is also a 
little disappointing. The energy levels and energy eigenfunctions are the 
same as that of a particle on a circle, except for the factor of $\alpha$. 
This is a consequence of the fact that, in the weak coupling limit 
(large $\gamma$), the putative torus degenerates into a limit circle, with 
the attendant vagueness associated with the winding in the $\theta$ direction. 
Correspondingly, the Mathieu functions degenerate into trigonometric 
functions. It may be tempting to think that the general solution (for 
arbitrary $\gamma$) will be given by Mathieu functions, with the boxed 
expressions in equations (35) and (37-38) being the next order corrections. 
The story, however, is slightly more complicated. Our penchant for boxing 
negligible pieces relates to this fact.    

\subsection{The Slightly-Thick-Torus Correction}

To the next order in $1/\gamma$, the correct expression for the potential 
is obtained by starting with equation (27), making a binomial expansion
of the denominator, and collecting terms up to order $1/\gamma^2$. This 
straightforward exercise, followed by the steps that lead up to equation 
(31), yields the so-called Hill-Whittaker equation \cite{ince}  
\begin{equation}
[{d^2\over dz^2} + \Theta_0 + 2\Theta_1 {\hbox{cos}}2z
+ 2\Theta_2 {\hbox{cos}}4z]\Sigma = 0
\end{equation}
with 
\begin{equation}
\Theta_0 = \lambda + {2\over 3\gamma^2},~~~~\Theta_1 = -{2\over 3\gamma},~~~~
\Theta_2 = -{1\over \gamma^2} 
\end{equation}
where now 
\begin{equation}
\lambda = {8MEa^2\over \hbar^2\alpha^2\gamma^2}
\end{equation}

Following Ince \cite{ince}, the most general solution of the Hill-Whittaker 
equation can be obtained along the same lines adopted for solving Mathieu's 
equation and yields the following energy eigenvalues
\begin{equation}
E_n = {\hbar^2\alpha^2\gamma^2\over 8Ma^2}\bigl\{ -4\nu^2 + 
(1 - {4\over 3\gamma}{\hbox{cos}}2\sigma - {2\over 3\gamma}{\hbox{sin}}2\sigma
-{7\over 9\gamma^2})\bigr\}
\end{equation}
The corresponding solutions are 
\begin{widetext}
\begin{equation}
\Sigma^{(n)} = e^{2inz/q}\bigl\{{\hbox{sin}}(z-\sigma ) + {2\gamma\over 3}
{\hbox {sin}}(3z - \sigma ) + ({1\over 108} - {4\gamma^2\over 9})
{\hbox{sin}}(5z - \sigma ) + {2\gamma\over 3} {\hbox{cos}}(3z - \sigma ) - 
{4\gamma^2\over 9} {\hbox{cos}}(5z - \sigma )
\bigr\}
\end{equation}
\end{widetext}
where $\sigma$ is the parameter introduced earlier. Once again multiplying 
by the factor $\chi = N\sqrt f$, expanding the denominator, retaining terms 
up to order $1/\gamma^2$ and, rewriting everything in terms of $\phi$ using 
$\alpha\phi = 2z$, gives the final expression for the eigenstate to be
\begin{widetext}
\begin{equation}
\begin{split}
\psi^{(n)}(\phi )= N&e^{in\alpha\phi/q}\times\\
&\bigl\{
-{4\gamma\over 9}[{\hbox{sin}}({5\alpha\phi\over 2}-\sigma ) 
+ {\hbox{cos}}({5\alpha\phi\over 2} -\sigma )]\\ 
&+{2\over 3}[{\hbox {sin}}({3\alpha\phi\over 2} 
- \sigma ) + {\hbox{cos}} ({3\alpha\phi\over 2} - \sigma ) - {2\over 3}
{\hbox{cos}} \alpha\phi \{{\hbox{sin}}({5\alpha\phi\over 2} -\sigma) - 
{\hbox{cos}}({5\alpha\phi\over 2} - \sigma )\}]\\ 
&+ {1\over\gamma} [{\hbox{sin}}({\alpha\phi\over 2} - \sigma ) + {1\over 108}
{\hbox{sin}}({5\alpha\phi\over 2} - \sigma ) 
+{2{\hbox{cos}}\alpha\phi\over 3} \{{\hbox{sin}}({3\alpha\phi\over 2} -\sigma ) 
 + {\hbox{cos}}({3\alpha\phi\over 2} -\sigma )\}\\
&\quad\quad\quad\qquad -{4{\hbox{cos}}^2\alpha\phi\over 9} \{{\hbox{sin}}
({5\alpha\phi\over 2}-\sigma ) 
 + {\hbox{cos}}({5\alpha\phi\over 2} -\sigma )\}]\\
&+{1\over\gamma^2} [{\hbox{cos}}\alpha\phi\{{\hbox{sin}}({\alpha\phi\over 2} - 
\sigma) + {1\over 108}{\hbox{sin}}({5\alpha\phi\over 2} -\sigma )\} 
+ {2{\hbox{cos}}^2\alpha\phi\over 3}\{{\hbox{sin}}({3\alpha\phi\over 2} -\sigma
) + {\hbox{cos}}({3\alpha\phi\over 2} - \sigma )\}\\ 
&\quad\quad\quad\qquad -{4{\hbox{cos}}^3\alpha\phi\over 9}\{{\hbox{sin}}
({5\alpha\phi\over 2} -\sigma ) + {\hbox{cos}}({5\alpha\phi\over 2} -\sigma 
)\}]\bigr\} 
\end{split}
\end{equation}
\end{widetext}
where $N$ is a normalization constant.

\subsection{The Result For An Arbitrarily Thick Torus}

For the sake of completion, we mention that this method can be systematically 
continued to arbitrary orders in $1/\gamma$. The corresponding equation 
satisfied by $\Sigma$ is the Hill equation given by 
\begin{equation}
[{d^2\over dz^2} + \Theta_0 + 2\sum_{r=1}^\infty\Theta_{2r} 
{\hbox{cos}}2rz]\Sigma = 0
\end{equation}
As in the earlier section, we follow Ince \cite{ince}, and try a 
general solution of the form 
\begin{equation}
\Sigma = e^{i\nu z}u
\end{equation}
where 
\begin{equation}
\nu = p_1(\sigma)\Theta_1 + p_2(\sigma)\Theta_2 + \cdots + q_1(\sigma)\Theta_1^2
+ q_2(\sigma)\Theta_2^2+\cdots +q_{12}\Theta_1\Theta_2 + q_{13}\Theta_1\Theta_3
 + q_{23}\Theta_2\Theta_3 + \cdots + r_1(\sigma)\Theta_1^3 +\cdots 
\end{equation}
and 
\begin{equation}
u = {\hbox{sin}}(z-\sigma) + A_1(z,\sigma)\Theta_1 + A_2(z,\sigma)\Theta_2
\cdots + B_1(z,\sigma)\Theta_1^2 + B_2(z,\sigma)\Theta_2^2 + \cdots +
B_{12}(z,\sigma)\Theta_1\Theta_2 + \cdots 
\end{equation}
with $\sigma$ being determined by the relation
\begin{equation}
\Theta_0 = 1 + \lambda_1(\sigma)\Theta_1 + \lambda_2(\sigma)\Theta_2 \cdots 
+\mu_1(\sigma)\Theta_1^2 + \mu_2(\sigma)\Theta_2^2 + \cdots +\mu_{12}(\sigma)
\Theta_1\Theta_2 \cdots +\nu_1(\sigma)\Theta_1^3 \cdots
\end{equation}
Substituting these expressions in equation (47) we can solve for the 
coefficients to any desired order, and hence obtain the corresponding
eigenvalues and eigenvectors. We don't pursue this exercise further
since it does not shed any further light on the solution to the problem.   

\section{Inequivalent Quantizations}

Let us briefly recapitulate the interesting consequences that arise if the 
particle which is constrained to move on a circle is charged, and if the 
circle encloses an infinitely long, infinitesimally thin, and impenetrable
solenoid carrying a uniform current. As is well-known, the wavefunction of 
the particle picks up a nontrivial phase factor which depends on the net 
flux enclosed by the trajectory of the particle as it goes around the circle. 
Thus the wavefunction is multi-valued, which is a manifestation of the 
nontrivial topology of the circle which, in turn, is a consequence of the 
fact that the path cannot be shrunk to a point in the presence of 
the impenetrable solenoid. Redefining the wavefunction such that it is 
single-valued modifies the Hamiltonian in such a way that the energy spectrum 
depends on the enclosed flux. Given that the corresponding Lagrangians,
with and without the flux, differ only by a total derivative term, the 
classical theory is unaltered; although different values of the flux yield 
different energy spectra, and hence inequivalent quantum theories.  
It is reasonable to expect similar features in the case of a charged
particle constrained to move along a knot. 

For the torus knot of interest, two independent magnetic fluxes can be 
introduced. The first is the usual magnetic field obtained by placing a
uniform current carrying, long, thin solenoid parallel to the $z$-axis 
and passing through the centre of the putative torus around which the knot 
winds. Let us denote the corresponding flux by $\Phi_S$. The second flux 
is obtained by a uniform poloidal current winding around the torus which 
produces a magnetic field which has a support only inside the torus, the 
so-called toroidal magnetic field. Let us denote this flux by $\Phi_T$. 

A particle constrained to move on a $(p,q)$ torus knot, starts at a point 
on the surface of the putative torus and returns to the initial point after 
completing one circuit of the knot; in the process winding around the 
solenoidal flux $p$ times and the toroidal flux $q$ times. The total 
flux enclosed is therefore: $\Phi = p\Phi_S + q\Phi_T$. Thus we have
the equation which highlights the multi-valued nature of the wavefunction
{\it viz.}
\begin{equation}
\psi (\eta_0, \theta + 2\pi q, \phi - 2\pi p) = {\hbox{exp}}(i\Phi)\psi 
(\eta_0, \theta, \phi) 
\end{equation}
Defining the single-valued wavefunction 
\begin{equation}
\tilde\psi (\eta_0, \theta , \phi ) = {\hbox{exp}}(-i{\Phi
\over 2p\pi}\phi )\psi (\eta_0, \theta, \phi) 
\end{equation}
and the corresponding Hamiltonian obtained by the transformation
\begin{equation}
\tilde H = {\hbox{exp}}(-i{\Phi \over 2p\pi}\phi ) H {\hbox{exp}}(i{\Phi 
\over 2p\pi}\phi ) 
\end{equation}
we see that the momentum operator in the Hamiltonian is shifted by 
${i\Phi\over 2p\pi}$, which leads to a corresponding shift in $\nu$
and hence the energy spectrum defined in equations (36) and (44).  
It is noteworthy that the phase picked up by the wavefunction of the 
particle, for a given $(p,q)$ knot, is a sum of two independent fluxes.
Thus the inequivalent quantizations are labelled by two parameters. 
   
\section{Conclusions and Outlook}

The classical and quantum mechanics of a particle constrained to 
move on a torus knot were studied. The results were compared and 
contrasted with the well-known results for a particle constrained 
to move on a circle. Defining the knot as a trajectory which winds
around a putative torus in a well-defined fashion, and using toroidal 
coordinates to parametrise the knot, makes it possible to rewrite the 
time-independent Schr\"odinger equation as a Hill equation which can
then be studied perturbatively in the thickness of the putative torus. 

Attributing a charge to the particle and introducing two independent 
magnetic fields having supports in physically disconnected, but 
topologically linked, regions, leads to a two-parameter family of
inequivalent quantizations of the particle moving on a knot.    

The model discussed in this paper has several features which are 
worth discussing further. First, it would be natural to 
study the model non-perturbatively {\it i.e.} using instanton 
methods made popular in \cite{cole}\cite{kittel}. Second, it would be 
interesting to generalise the treatment to more than one particle moving on 
the knot. The non-trivial phase factor can then be related to 
exotic quantum statistics of the anyonic type. It would also be 
interesting to construct coherent states and study algebras of 
quantum observables associated with a particle on a knot. All these 
problems have natural analogues for the corresponding, but 
much simpler, example of a particle constrained to move on 
a circle.

\begin{acknowledgements}

I thank G. Krishnaswami and A. Laddha for discussions. This work is 
partially funded by a grant from Infosys Foundation.  

\end{acknowledgements}

\end{document}